\documentclass[a4paper,12pt]{article}
\usepackage[top=3cm,bottom=2cm,left=3cm,right=3cm]{geometry}
\usepackage{amsmath}
\usepackage{graphicx}
\usepackage{bm}
\usepackage{lineno}
\usepackage{subfigure}
\usepackage{url}
\usepackage{authblk}
\usepackage[colorlinks,unicode=true]{hyperref}
\usepackage{ulem}
\usepackage{float}

\begin{document}
\title{PHASE-RESOLVED GAMMA-RAY SPECTROSCOPY OF THE CRAB PULSAR OBSERVED BY POLAR}

\author[abc]{Han-Cheng Li\thanks{Corresponding author at: Key Laboratory of Particle Astrophysics, Institute of High Energy Physics, Chinese Academy of Sciences, Beijing 100049, China.\newline E-mail address: Hancheng.Li@ihep.ac.cn}}
\author[d]{Neal Gauvin}
\author[a]{Ming-Yu Ge}
\author[e]{Wojtek Hajdas}
\author[c]{Merlin Kole}
\author[a]{Zheng-Heng Li}
\author[d]{Nicolas Produit}
\author[ab]{Li-Ming Song}
\author[ab]{Jian-Chao Sun}
\author[f]{Jacek Szabelski}
\author[f]{Teresa Tymieniecka}
\author[ab]{Yuan-Hao Wang}
\author[a]{Bo-Bing Wu}
\author[c]{Xin Wu}
\author[a]{Shao-Lin Xiong}
\author[ab]{Shuang-Nan Zhang}
\author[a]{Yong-Jie Zhang}
\author[a]{Shi-Jie Zheng}

\affil[a]{Key Laboratory of Particle Astrophysics, Institute of High Energy Physics, Chinese Academy of Sciences, Beijing 100049, China}
\affil[b]{University of Chinese Academy of Sciences, Beijing 100049, China}
\affil[c]{Department of Nuclear and Particle Physics, University of Geneva, 24 Quai Ernest-Ansermet, 1205 Geneva, Switzerland}
\affil[d]{University of Geneva, Geneva Observatory, ISDC, 16, Chemin d'Ecogia, 1290 Versoix Switzerland}
\affil[e]{Paul Scherrer Institut, 5232, Villigen, Switzerland}
\affil[f]{National Centre for Nuclear Research, ul. A. Soltana 7, 05-400 Otwock, Swierk, Poland}
\maketitle


\newpage

\begin{abstract}
The POLAR detector is a space based Gamma-Ray Burst (GRB) polarimeter sensitive in the 15-500 keV energy range. Apart from its main scientific goal as a Gamma-Ray Burst polarimeter it is also able to detect photons from pulsars in orbit. By using the six-months in-orbit observation data, significant pulsation from the PSR B0531+21 (Crab pulsar) was obtained. In this work, we present the precise timing analysis of the Crab pulsar, together with a phase-resolved spectroscopic study using a joint-fitting method adapted for wide field of view instruments like POLAR. By using single power law fitting over the pulsed phase, we obtained spectral indices ranging from 1.718 to 2.315, and confirmed the spectral evolution in a reverse S shape which is homogenous with results from other missions over broadband. We will also show, based on the POLAR in-orbit performance and Geant4 Monte-Carlo simulation, the inferred capabilities of POLAR-2, the proposed follow-up mission of POLAR on board the China Space Station (CSS), for pulsars studies.
\end{abstract}

\providecommand{\keywords}[1]
{
  \small	
  \textbf{Key words:} #1
}
\keywords{POLAR; gamma-ray; neutron star; pulsars; PSR B0531+21 (Crab pulsar); spectroscopy}

\section{Introduction}
\label{sec:intro}
The Crab Nebula is a so-called pulsar wind nebula (PWN)~\cite{Weiler(1978)}, associated with the supernova explosion which happened in 1054 AD and recorded by Chinese astronomers. In its very center lies the Crab pulsar as an energetic engine. The Crab pulsar has a stable spin period of $\sim$\,33 ms with derivative of $\sim$\,$4.2\times{10^{-13}}$ s/s, proximal distance of $\sim$\,2 kpc, and bright radiation over almost the full electromagnetic spectrum from radio ($\sim$\,$10^{-5}$ eV) to high energy $\gamma$-rays ($\sim$\,$10^{12}$ eV)~\cite{Fermi_LAT_Crab(2010), MAGIC_Crab(2016)}. These features make it one of the best studied celestial objects, and thus a common calibration source for astronomical missions~\cite{HXMT_Crab(2019)}. 
The Crab pulsar shows a double-pulse structure, with the main pulse (MP) and the inter pulse (IP) separated by the pulse phase of $\sim$\,$144^{\circ}$ (0.4 in 1) approximately aligned in absolute phase in every energy band~\cite[and references therein]{Eikenberry(1997),Kuiper(2001),Rots(1998),Molkov(2010),Ge(2012),Eilek(2016)}. The studies for the origin and structure of the particle acceleration region of pulsars (PSRs) over the past fifty years have evolved with multiple theoretical models (see review paper~\cite[and references therein]{Harding(2017)}) that suggesting different radiation regions and mechanisms. From observation, the pulsed emissions of PSRs can be selected to study the physical properties, especially the phase-resolved spectroscopy plays a key role in understanding radiation characteristics of regions under different rotating phases, thereby providing important constraint on theoretical models. The first phase resolved X-ray spectroscopy of the Crab pulsar, performed by Pravdo and Serlemitsos~\cite{Pravdo(1981)}, suggested that the evolution of the spectral index versus the pulse phase was acting like a reverse S shape. This was later validated and improved by other measurements. More recently, missions including RXTE~\cite{Ge(2012)}, INTEGRAL~\cite{Integral_Crab(2006)}, Fermi LAT~\cite{Fermi_LAT_Crab(2010)} and HXMT~\cite{HXMT_Crab(2019)} produced new results with better statistics and finer phase binning over different energy ranges.

POLAR is a dedicated Gamma-Ray Burst (GRB) polarimeter which took data on the second Chinese space laboratory Tiangong-2 (TG-2) from September 2016 to April 2017. In order to optimize the instrument for GRB polarization measurements, it has a wide Field of View (FoV) of about 2$\pi$ steradian and an effective area of approximately $400\,\mathrm{cm}^2$ at 300 keV~\cite{Produit(2018)}. These features make it one of the most sensitive instruments for GRB detections in its energy range (15-500 keV). Installed on board the TG-2, with an orbit inclination of $42^\circ$ and always pointing towards the zenith of the instrument, POLAR scanned a large fraction of the sky every orbit. This, together with its high sensitivity, makes it capable of almost continuously monitoring point sources such as PSRs. More details of the POLAR detectors construction can be reffered to \cite{Produit(2018),Produit(2005)}, while the descriptions of the instrument on-ground and in-orbit performance and simulation software the can be referred to~\cite{Simulation(2017)} and~\cite{Calibration(2018)}.

Apart from studying a method to perform detailed spectral measurements for the Crab pulsar using a non-pointing wide FoV instrument, the study presented here also serves as a calibration before proceeding with the polarization studies of the Crab pulsar using POLAR data. As the POLAR instrument is not dedicated for persistent sources, such a study is complicated by a low signal to background ratio (SNR). This however, can be compensated by the large exposure to the source. And perhaps more challenging obstacle is the detailed calibration required to perform polarization measurements. As several past polarization measurements are not trusted by the community due to a lack of calibration, see for example \cite{McConnell(2017)}, we decided to first carefully check the accuracy of the POLAR calibration for all incoming photon angles.

In this paper, we present the Crab pulsar detection by POLAR with its precise timing and phase-resolved spectral analysis, both of which are also expected to serve as calibrations. In Section~\ref{sec:reduction} we show the data and reduction process. The timing of Crab pulsar is presented in Section~\ref{sec:timing}. In Section~\ref{sec:fitting} we perform a phase-resolved spectral analysis with the method introduced in. The sensitivity of POLAR-2 to the Crab pulsar is discussed in Section~\ref{sec:polar2}. Finally, we present the main conclusions of this paper in Section~\ref{sec:discussion}.

\section{Observation and Data Reduction}
\label{sec:reduction}
POLAR captured billions of photons during its six-months operation, only a fraction of them originate from PSRs and were detected when they entered POLAR's FoV. The absolute time of each arrival photon is computed by using GPS, and we have position of POLAR and the attitude from TG-2 telemetry in every second with great accuracy. The direction of incident light from the Crab pulsar is then estimated in POLAR's local coordinate. The Crab is visible in POLAR when the incident zenith angle is smaller than $\sim$\,$102^{\circ}$~\cite{Hancheng(2017)}. Since POLAR has no pointing control system by itself, the pointing of Z-aixs of POLAR detector is rotating at a speed of 4 arcmin per second. The incident angle of the Crab pulsar is thus continuously changing with time. Additionally as the instrument has a wide FoV, rough capabilities in source localization and restricted particle identification capability, the SNR for this study is thus relatively low. This is however compensated by the large exposure time to the Crab pulsar. Event selection and data reduction applied here are based on the methods described in~\cite{Calibration(2018)}. During this process the pedestals are subtracted, common noise is removed and cross talk and energy calibration are applied for each event. Additionally cosmic ray events and triggers induced by anomalies in the electronics are removed using the methods described in detail in~\cite{Calibration(2018)}. For this study additionally we binned the data into one second light curve and we excluded intervals with high background, thus defining Good Time Interval (GTI) which mainly include periods where the instrument was close to the South Atlantic Anomaly (SAA), GRBs and solar flares, etc. To get the Observation Datasets (ODs) for the Crab pulsar, we considered that: 

1) we use the HEALPIX~\cite{Healpix(2005)} to divide POLAR's local sky into 768 healpix bins by setting $N_{side}$, a resolution parameter from HEALPIX, equal to 8. In this paper, we use RING numbering schedule, which means that pixels from 0 to 767 are result of moving down from the zenith to the anti-zenith of POLAR and anticlockwise along its X-Y plane. The incident angle of the Crab pulsar was used to determine which healpix bin it is in, therefore to classify the healpix datasets for it. But considered the conditional visibility of the Crab pulsar, we kept healpix bins from 0 to 399, so we have 400 healpix datasets for the Crab pulsar; 

2) It should also be clarified here that POLAR ran two different observation modes during its mission with different trigger strategies: single-mode (SM) and double-mode (DM). During DM mode events usable for polarization studies were stored, meaning only events where at least 2 of the 1600 channels had a signal above trigger level. Additionally only 1 in every 1000 events where only 1 channel was above trigger level was stored, a process referred to as pre-scaling which was used to reduce the data size. During SM mode pre-scaling was removed and as a result all single multiplicity trigger events were stored, thereby greatly increasing the sensitivity of the instrument. 

Finally, after selection, events were then assigned to different healpix bin and observation mode. As a result originally we have 800 ODs for the Crab pulsar including 400 healpix bins in SM and 400 healpix bins in DM. The instrument response matrices were generated for all 800 ODs by using the simulation package described in~\cite{Simulation(2017)} and the input parameters from calibration~\cite{Calibration(2018)}. We calculated the significance of the Crab pulsation ($S_{psr}^i$) for OD $i$ (i from 1 to 800) by using:

 \begin{equation}\label{eq:Sign}
S_{psr}^i = \frac{N_{src}^i-w\times{N_{bkg}^i}}{\sqrt{N_{src}^i+w^{2}\times{N_{bkg}^i}}},
\end{equation}

\begin{figure}[!htbp]
\centering
\includegraphics[width= 0.86\textwidth, height = 6cm]{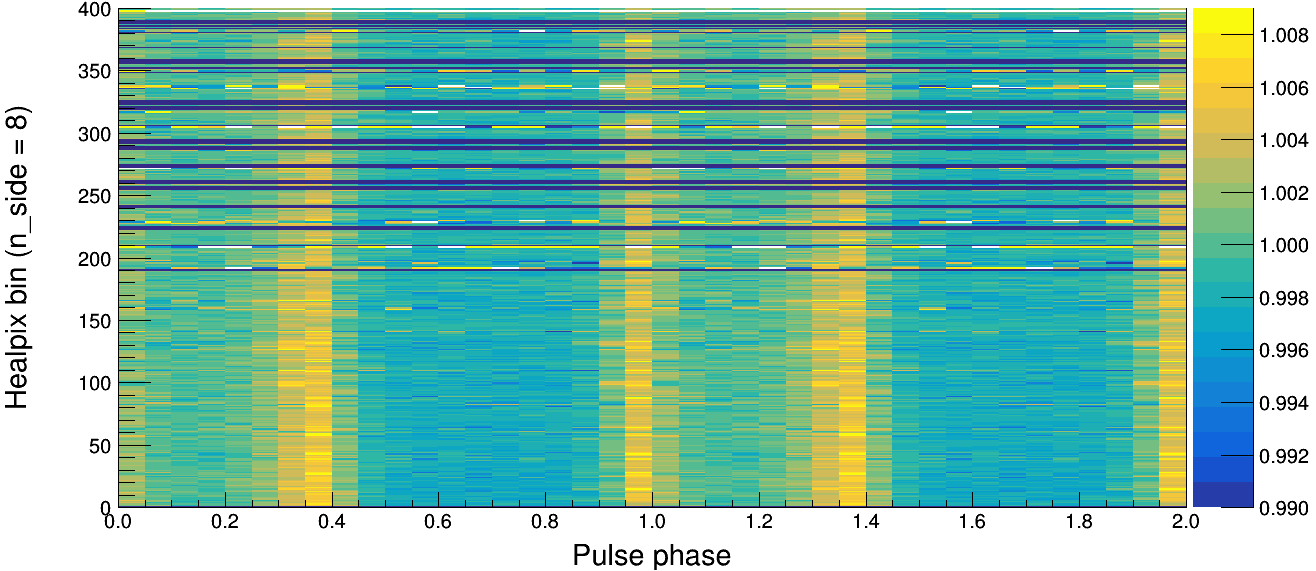}
\caption{\textbf{The pulse profile as a function of healpix bin.} The normalized pulsar profiles (along the x-axis) as a function of healpix bin (y-axis). These empty bins along the y-axis are a result of a lack of data due to not having any exposure to the Crab for these specific angles. It can be seen that in each healpix bin the profiles are aligned with one another.}\label{fig:PulsePheal}
\end{figure}

\begin{figure}[!htbp]
\centering
\includegraphics[width= 0.86\textwidth, height = 12cm]{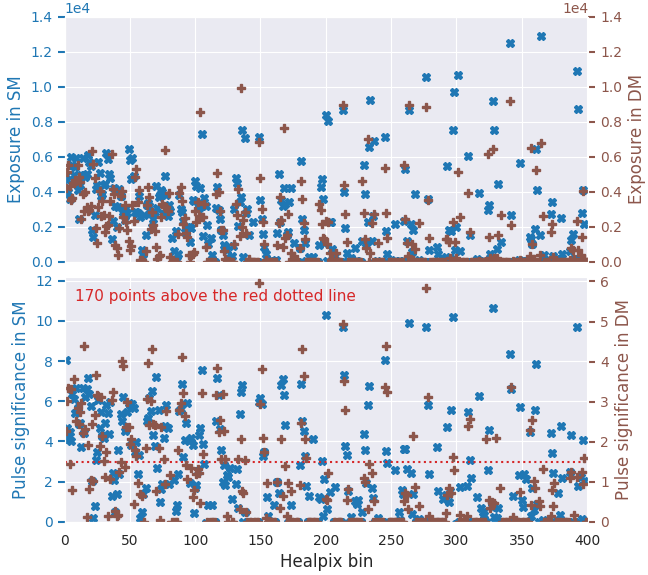}
\caption{\textbf{Pulse significance and exposure time vs healpix bin.} The top panel shows the exposure time of the Crab pulsar as a function of the healpix bin both for single mode (x marker) and double mode (+ marker); The bottom panel shows the corresponding pulse significance as a function of healpix bin both in single mode (x marker) and double mode (+ marker). The red dotted line is the 3 sigma cut line used to select a sub-sample of ODs. A total of 170 points are selected. Note that we are in RING numbering schedule as defined by HEALPIX, this implies that healpix bin 0 is at detector zenith.}\label{fig:exposure_sign}
\end{figure}

Where $w$ is the ratio of the length of pulsed phase interval to the length of background phase interval, for the Crab pulsar $w$ is 4.0 since the background phase interval is 0.6-0.8, which is equal to that used in~\cite{HXMT_Crab(2019),Integral_Crab(2006)}, and the pulsed phase interval is the rest parts. $N_{src}^i$ and $N_{bkg}^i$ are the total counts of pulsed phase interval and background phase interval of OD $i$, respectively. Then, for the spectral analysis, we cut $S_{psr}^i$ at 3 $\sigma$ (red dotted line in the bottom panel of Figure~\ref{fig:exposure_sign}) to select a sub-sample of ODs. In Figure~\ref{fig:PulsePheal}, we show the pulse profile as a function of the healpix bin. The exposure time and pulse significance as a function of healpix bin in SM and DM is shown in Figure~\ref{fig:exposure_sign} top and bottom panels respectively. The rest sub-sample of 170 ODs for Crab pulsar after cutting pulse significance at 3 $\sigma$ is listed in Table~\ref{table:ObservationODs}.

\section{Timing}
\label{sec:timing}
The time of arrival (ToA) of each photon measured by POLAR was corrected to the corresponding time in the Solar System Barycenter frame by using ephemerides of DE405, all the related time delays were take into account. We have studied the timing properties of the Crab pulsar with the POLAR data to get accurate timing parameters~\cite{Hancheng(2017)}, they were validated against Fermi. Then the pulse phase $\phi_i$ of photon $i$ was calculated by:

 \begin{equation}\label{eq:Phi}
\phi_i = f_{0}(t_{i}-t_{0}) + \frac{1}{2}f_{1}(t_{i}-t_{0})^2 + \frac{1}{6}f_{2}(t_{i}-t_{0})^3 + \frac{1}{24}f_{3}(t_{i}-t_{0})^4 + \cdots,
\end{equation}

where $t_i$ is the corrected ToA of photon $i$, $t_0$ is the reference time zero, $f_0$ is the periodic frequency, and $f_{1}$, $f_{2}$ and $f_{3}$ are derivative, second derivative and third derivative of the frequency at $t_0$, respectively. Timing parameters are taken from~\cite{Hancheng(2017)}. 

\begin{figure}[!htbp]
\centering
\includegraphics[width= 0.86\textwidth, height = 8cm]{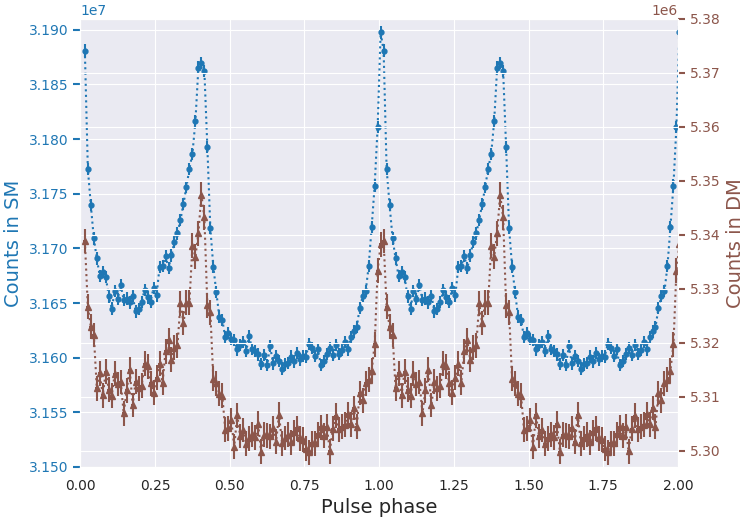}
\caption{\textbf{The pulse profile of SM and DM.} The total pulse profile accumulated from singles mode (point marker) and double mode (triangle marker). Since the trigger strategies are different between SM and DM the sensitivity to the Crab pulsar in two modes differs significantly.}\label{fig:pulse_sd}
\end{figure}

\begin{figure}[!htbp]
\centering
\includegraphics[width= 0.86\textwidth, height = 8cm]{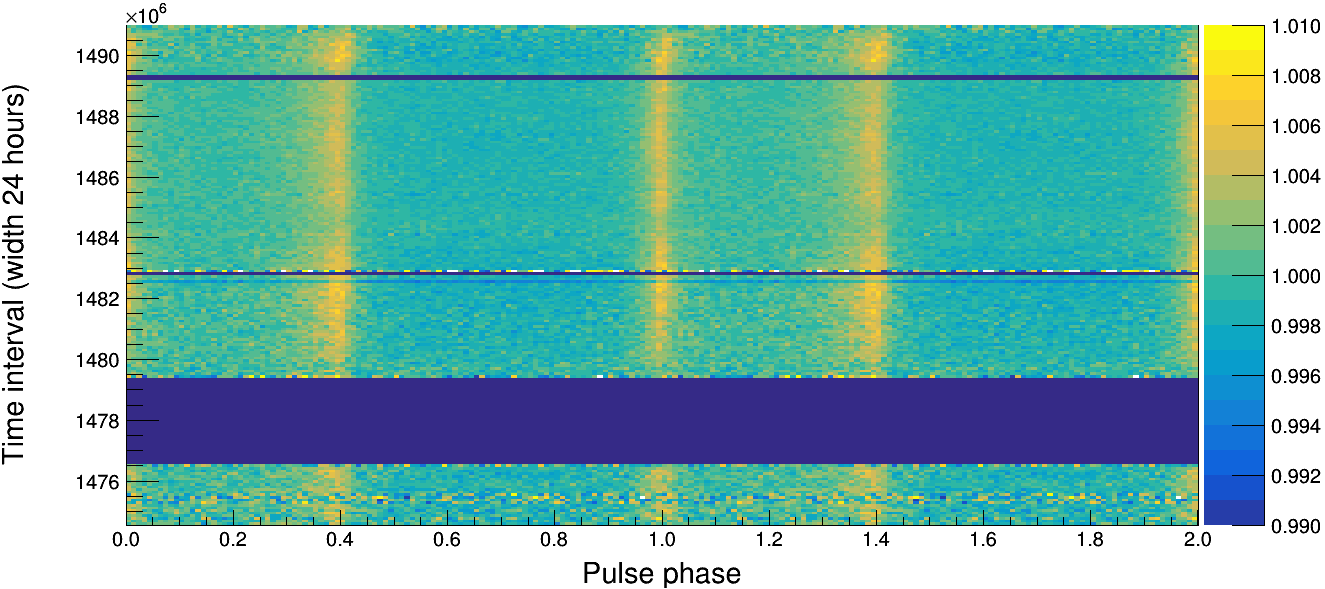}
\caption{\textbf{The pulse profile as a function of unix time.} The normalized profiles with background subtraction observed as a function of unix time. The binning along the y-axis corresponds to 1 day per bin. These profiles can be seen to be co-aligned. The empty bins along y-axis are due to the instrument being turned off during these days. Additionally, there were several days the instrument was either in calibration mode or was switched off for part of the day, resulting in less statistics.}\label{fig:waterfall}
\end{figure}

After phase folding, the total pulse profile of the Crab pulsar using all 400 healpix bins both in SM and DM are obtained, as shown in Figure~\ref{fig:pulse_sd}. A profile was produced using the data for each day of the mission. These profiles are co-aligned during the full mission with the same phase as illustrated in Figure~\ref{fig:waterfall}. These results confirm that POLAR has detected the pulsed photons from the Crab pulsar and that the timing of the instrument is found to be stable during operation.

\section{Phase-resolved spectral fitting}
\label{sec:fitting}
For each OD, we produce phase-resolved spectra for the Crab pulsar. In this study we considered the 15-500 keV measured energy range, where 15 keV is the average threshold of each bar in the POLAR detector. We appropriately divide the pulsed phase interval into finer phase bins listed in the first column of Table~\ref{table:resultstable} which includes 40 phase bins in total. The phase binning was based on that used by~\cite{HXMT_Crab(2019),Integral_Crab(2006)} and corrected for the reduced statistics in this study compared to those in~\cite{HXMT_Crab(2019),Integral_Crab(2006)}. For each phase bin we extract a spectrum by selecting photons in the phase range of this bin. The background spectrum is extracted in phase range from 0.6 to 0.8, which is equal to that used in~\cite{HXMT_Crab(2019),Integral_Crab(2006)}. The exposure time $E_{i}$ of any continuous phase range $i$ can be calculated by:

\begin{equation}\label{eq:exposure}
E_{i} = \frac{len_{i}}{1.0}\times{E_{all}}
\end{equation}

Where $len_{i}$ is the phase length of interval $i$, $E_{all}$ is the total exposure time of one OD. Based on this, the exposure time for each phase bin and for background can be calculated. The statistical error of each spectral channel is assumed to be Poisson distributed. This process was applied to the 170 ODs listed in Table~\ref{table:ObservationODs}. As a result for each phase bin we have 170 pulsed spectra and 170 corresponding background spectra with their exposure time, and from simulation we get 170 corresponding instrumental response matrices.

We used XSPEC (\textbf{version: 12.10.1})~\cite{XSPEC12} to fit our spectra using a single power law model similar to one used in~\cite{HXMT_Crab(2019),Integral_Crab(2006)}. We adapted all data to the OGIP FITS standard~\cite{OGIP}. XSPEC normalizes the background spectrum in the following way~\footnote{ \url{https://heasarc.gsfc.nasa.gov/docs/asca/abc\_backscal.html}}:

\begin{equation}\label{eq:subtraction}
net\_spectrum =
\frac{src\_counts}{src\_exposure} - 
  \frac{bkg\_counts}{bkg\_exposure}
    \times{\frac{src\_BACKSCAL}{bkg\_BACKSCAL}}
\end{equation}

Where $src\_counts$ is a spectrum of each phase bin, $bkg\_counts$ is its corresponding background spectrum. Here we set $src\_BACKSCAL$ and $bkg\_BACKSCAL$ to 1.0 for our observations. The error propagation is considered by XSPEC automatically.

\begin{figure}[!htbp]
\centering
\includegraphics[width= 0.86\textwidth, height = 7.5cm]{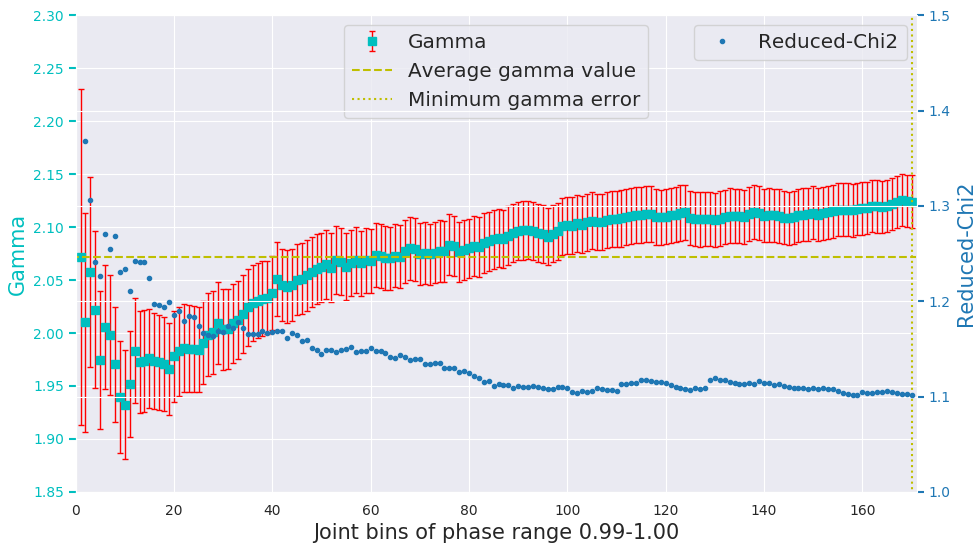}
\caption{\textbf{The Gamma/Error vs Joint-Fitting bins.} The fitting results as a function of number of selected ODs. The cyan points indicate the best fitted $\Gamma$ values. The red error bar is its 1 $\sigma$ error. The blue points are the reduced $\chi^2$ values of each fit. These can be seen to approach $\sim$1.1 (in this case) when adding more OD. The errors can be seen to decrease. The minimum gamma error is the last point which indicating that the fit converges when increasing the data. The best fitted $\Gamma$ values are converging with more ODs joining in, the average $\Gamma$ value of them equals to the best fitted vale of the first bin.}\label{fig:JointBins}
\end{figure}

Since POLAR has multiple ODs, for each phase bin, we need to fit 170 background-subtracted spectra simultaneously. In order to check joint-fitting method based on XSPEC, we performed a series of fitting by adding joint ODs one by one following the order from Table~\ref{table:ObservationODs}. We use phase bin in range of 0.99-1.00 (in the MP) as an example, the patterns of other phase bins are similar with this one. The best fitted power law spectral index $\Gamma$ and its 1 $\sigma$ error as a function of joint OD is shown in Figure~\ref{fig:JointBins}. We found that the 1 $\sigma$ error values are decreasing, and the best fitted $\Gamma$ values are converging with more ODs joining in as expected. In addition, the reduced $\chi^2$ values are approaching to $\sim$1.1 (NDF=16318) from the joint-fitting of this phase bin. This result suggests that this joint-fitting method is statistically reasonable. Therefore for each phase bin, we used all 170 ODs of the sub-sample to fit.
 
\begin{figure}[!htbp]
\centering
\includegraphics[width= 0.86\textwidth, height = 7.5cm]{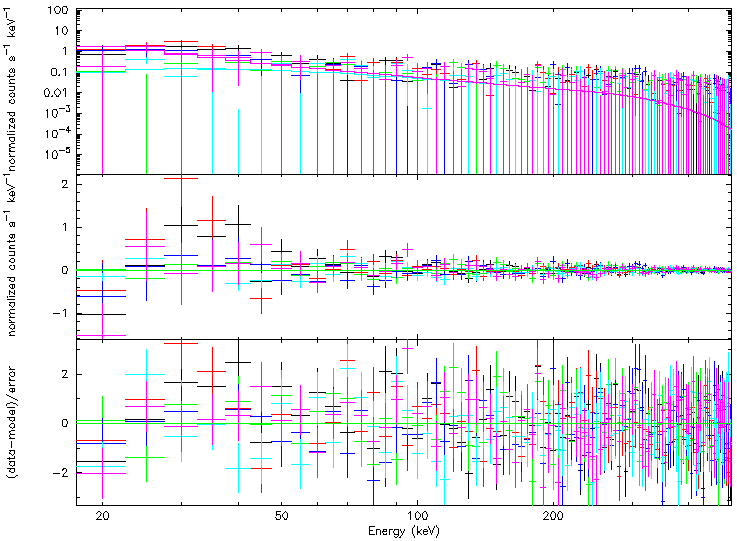}
\caption{\textbf{Spectral energy distribution of the Crab pulsar in phase range 0.99-1.00} We use the spectra of the first 5 ODs (in order to prevent the image from being too crowded). The top panel shows the spectral energy distribution, different colors represent different ODs, the magenta continuous line is the best fit with a single power law. The middle and the bottom panels are the residuals distribution in counts/s/keV and in terms of sigmas with error bars of size one respectively. }\label{fig:spectrumeg}
\end{figure}

\begin{figure}[!htbp]
\centering
\includegraphics[width= 0.9\textwidth, height = 8cm]{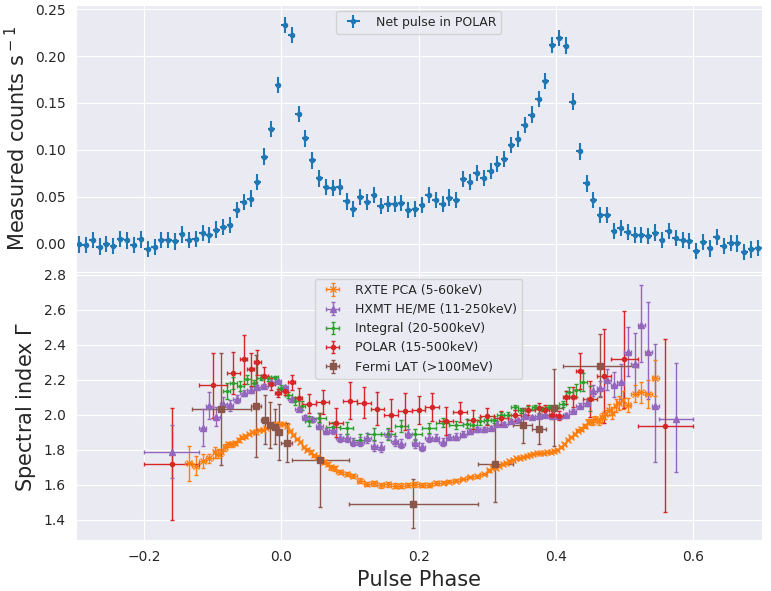}
\caption{\textbf{The spectral index vs pulse phase.} The top panel shows the background subtracted pulse counts rate accumulated from 170 ODs with a total exposure of 1432349.0 seconds. Here the 0.00 in the y-axis of the top panel is actually 26.88 counts/s for the mean background counts rate before subtraction. The bottom panel shows the best fitted $\Gamma$ values and its 1 $\sigma$ errors as a function of pulse phase.}\label{fig:spectra}
\end{figure}

Finally, we performed joint-fitting for a total of 40 phase bins by using 170 ODs with reduced $\chi^2$ from 0.995 to 1.102, the results are listed in Table~\ref{table:resultstable}. We use the spectra of phase bin in range of 0.99-1.00 to visualize the fit results, where only the first 5 ODs in Table~\ref{table:ObservationODs} were used to illustrate (in order to prevent the image from being too crowded, we didn't show the one used all 170 ODs). The result is shown in Figure~\ref{fig:spectrumeg}. We plot the best fitted $\Gamma$ values and their 1 $\sigma$ error as a function of pulse phase in Figure~\ref{fig:spectra}, where we also put results from RXTE~\cite{Ge(2012)}, HXMT~\cite{HXMT_Crab(2019)}, Integral~\cite{Integral_Crab(2006)} and Fermi LAT~\cite{Fermi_LAT_Crab(2010)} for comparison. In our result, points near the edge of background phase interval and in the bridge phase range have large errors due to a lack of statistics, therefore they were fluctuating comparing to results of other missions. Our spectral index distribution can be observed to match those from instruments which took data in the same energy range, they all acted like a inverse S shape, thereby confirming both the results of such instruments, as well as the calibration of the POLAR detector. As for results from RXTE~\cite{Ge(2012)} and Fermi LAT~\cite{Fermi_LAT_Crab(2010)}, whose energy band are lower and higher than ours respectively, their evolution pattern are homogeneous with ours, but deviations among us indicate that spectra both from higher and lower energy band are harder than that of ours. The present results may appeal to future missions with high sensitivity over broadband for joint analysis.

\section{POLAR-2 detectivity for the Crab pulsar}
\label{sec:polar2}
POLAR-2 is successor of POLAR, which has been selected to be installed on board the China Space Station (CSS) officially with a launch schedule in 2024. Based on the preliminary design of POLAR-2 which has 4 times the number of the detector modules of POLAR, we performed Monte-Carlo simulations with Geant4 to calculate the effective area as a function of energy. Then from the convolution of the effective area and the Crab pulsar spectrum, we found the amount of measured photons of POLAR-2 is approximately 10 times that of POLAR in the 15-500 keV energy range. We considered that the amount of background photons is in direct proportion to the geometrical area of the instrument, therefore for POLAR-2 we conservatively take 4 times of the background level to POLAR. Detailed background simulations are however required in the future to provide a more accurate number for this.

\begin{figure}[!htbp]
\centering
\includegraphics[width= 0.9\textwidth, height = 8cm]{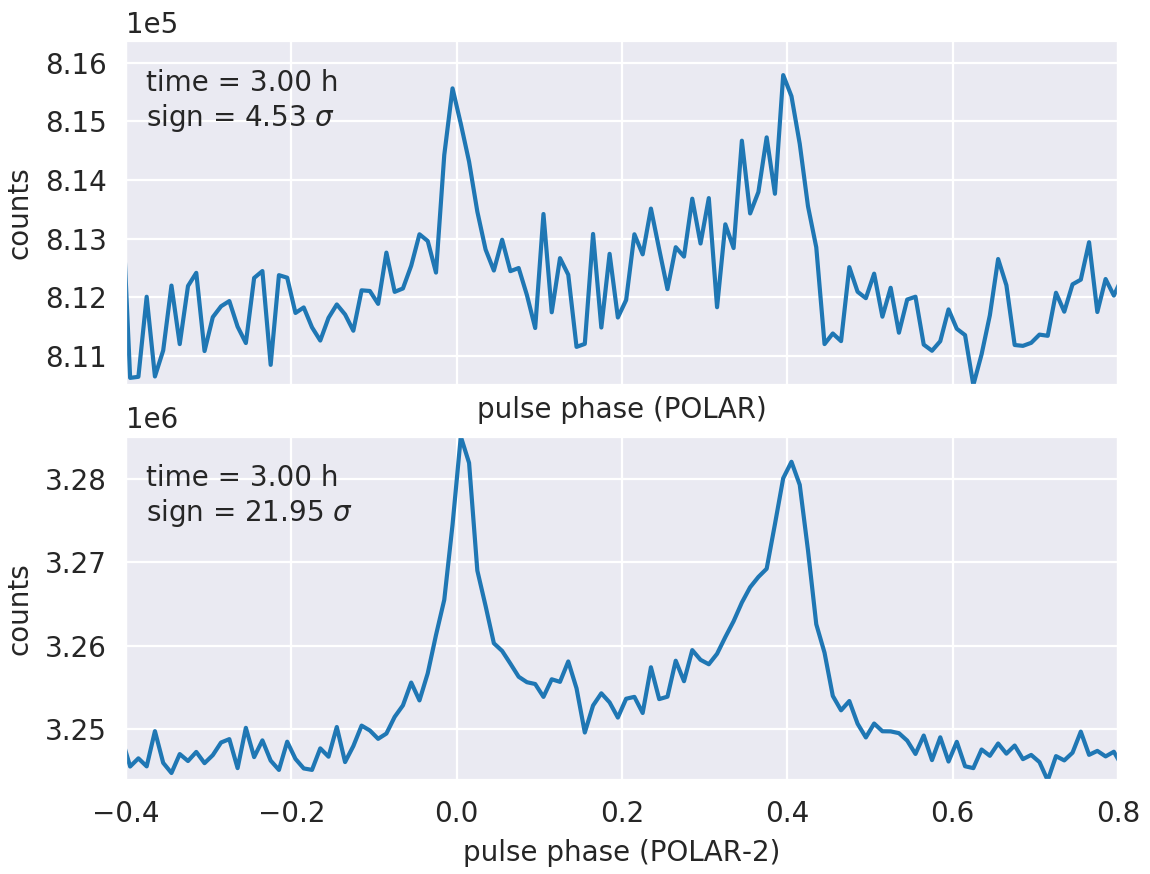}
\caption{\textbf{Three-hour observation of POLAR and POLAR-2.} The top panel shows the pulse profile of the Crab using three hours of observation of POLAR. The bottom panel shows the pulsar profile achieved using a three hour observation of POLAR-2.}\label{fig:threehours}
\end{figure}

By using the observed pulse count rate distribution shown in the top panel of Figure~\ref{fig:spectra} and the mean background count rate of 26.88 counts/s, we sample both pulse and background observations by factor of the given exposure time divided by 1432349.0 seconds (the total exposure time for the Crab pulsar during the POLAR mission), the statistical error of background sample is assumed as Poisson distributed. Adding pulse and background observations together, we obtain the sampled total observation of the Crab pulsar for a given exposure time, and the significance of it can be calculated by Eq~\eqref{eq:Sign}. We generate a movie of the estimated pulse profiles as a function of exposure time of Crab pulsar both for POLAR and POLAR-2~\footnote{ \url{https://www.astro.unige.ch/polar/supplement_1}\label{ft:video}}. In Figure~\ref{fig:threehours} we select the typical three-hour observation to show the difference of pulse profiles between POLAR and POLAR-2. In POLAR three-hour observation just exceeds the cut value of the pulse significance at 3 $\sigma$ we used in this study. While using the same exposure time for POLAR-2, we can obtain a much more significant pulsation.

\begin{figure}[!htbp]
\centering
\includegraphics[width= 0.86\textwidth, height = 7cm]{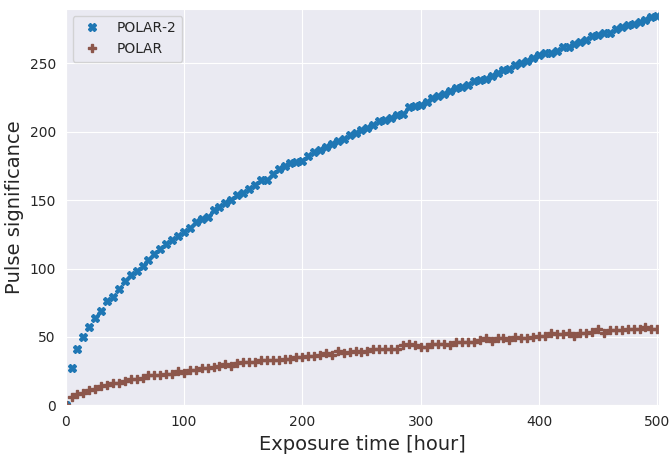}
\caption{\textbf{Pulse significance as a function of exposure time for POLAR and POLAR-2.} The averaged pulse significance of POLAR-2 (x marker) can be seen to be about 5 times that of POLAR (+ marker). The full video to show how the Crab pulse significance increase with time in POLAR and in POLAR-2 can be found in Footnote$^{~\ref{ft:video}}$}\label{fig:signPOLAR2}
\end{figure}

We calculated the pulse significance from the estimated pulse profiles as a function of exposure time of the Crab pulsar both for POLAR and POLAR-2. In Figure~\ref{fig:signPOLAR2} we show 500 hours distribution to reveal the capacity variance of pulsar detection between POLAR and POLAR-2. In this paper we used $\sim$400 hours exposure time for the Crab pulsar from POLAR to perform the spectral fitting, the total significance is 58.14. However using an exposure time of 33 hours for the Crab pulsar, POLAR-2 can exceed this pulse significance and be able to do equivalent spectral fitting presented above. Using the about 2 years of POLAR-2 data a significantly more detailed spectral analysis of the Crab pulsar, even of other PSRs, can therefore be performed. As far as now POLAR confirmed 2 PSRs detection (PSR B0531+21 and PSR B1509-58), and could potentially search for one more according to the flux of known PSRs~\cite{knownflux}. While by extrapolating these results presented here, POLAR-2 is promising to detect $\sim$10 PSRs.

\section{Conclusions}
\label{sec:discussion}
The results of timing of Crab pulsar suggest a good performance of POLAR, and phase-resolved spectroscopy of Crab pulsar indicates a proper calibration of the instrument. A wide FoV detector without pointing ability is able to do a spectral fitting for persistent celestial sources using the method presented here. The pattern of the evolution of spectral indices as a function of the pulse phase is consistent with the results from other missions. The results indicate that not only the full instrument response is modelled well but also that this response as a function of incoming angle is accurately modelled in the POLAR simulations software. The results presented here indicate that polarization studies of the Crab pulsar, which require accurate instrumental responses, can be further investigated. Additionally the inferred detectivity of POLAR-2 for PSRs will be improved significantly than POLAR.

\section*{Acknowledgments}
\label{sec:ackonwledgment}

We gratefully acknowledge the financial support from the Joint Research Fund in Astronomy under the cooperative agreement between the National Natural Science Foundation of China and the Chinese Academy of Sciences (Grant No. U1631242), the National Natural Science Foundation of China (Grant No. 11503028, 11403028), the Strategic Priority Research Program of the Chinese Academy of Sciences (Grant No. XDB23040400), the National Key R\&D Program of China (2016YFA0400800), the National Natural Science Foundation of China (Grants U1838201, U1838202 and U1838104), the Youth Innovation Promotion Association of Chinese Academy of Sciences (Grant No. 2014009), the Xie Jialin Foundation of the Institute of High Energy Phsyics, Chinese Academy of Sciences (Grant No. 542019IHEPZZBS10210), the Swiss Space Office of the State Secretariat for Education, Research and Innovation (ESA PRODEX Programme), and the National Science Center of Poland (Grant No. 2015/17/N/ST9/03556). This work is also supported by the University of Chinese Academy of Sciences (UCAS) Joint PhD Training Program (UCAS 2018 awardee No.4). We furthermore thank Ling-Song Ge of ISDC, Geneva, Switzerland, for beneficial discussions about the OGIP FITS standard and XSPEC. We thank the reviewers for their valuable comments that helping us improving this manuscript.

\bibliographystyle{unsrt}  

\newpage

\begin{table}[!htbp]
\caption{Sub-sample of the observations for Crab pulsar \protect\footnotemark}\label{table:ObservationODs}
\centering
\begin{tabular}{rcccc}
\hline\hline
No.OD	&	theta/$^{\circ}$	&	phi/$^{\circ}$	&	significance	&	exposure/s \\ 
\hline
327-SM	&	80.4059	&	264.375	&	10.6447	&	9192.58 \\ 
199-SM	&	60	&	264.375	&	10.2782	&	8383.44 \\ 
296-SM	&	75.5225	&	270	&	10.194	&	7528.12 \\ 
263-SM	&	70.5288	&	264.375	&	9.90218	&	8683.37 \\ 
276-SM	&	75.5225	&	45	&	9.71786	&	10576.8 \\ 
391-SM	&	90	&	264.375	&	9.71184	&	10921.6 \\ 
212-SM	&	65.3757	&	45	&	9.69444	&	8683.46 \\ 
340-SM	&	85.2198	&	45	&	8.3464	&	12501.8 \\ 
244-SM	&	70.5288	&	50.625	&	8.05555	&	7158.66 \\ 
0-SM	&	5.85027	&	45	&	8.04245	&	4951.68 \\ 
360-SM	&	85.2198	&	270	&	7.856	&	6467.46 \\ 
104-SM	&	41.8588	&	263.571	&	7.57692	&	7298.97 \\ 
213-SM	&	65.3757	&	56.25	&	7.31948	&	3997.23 \\ 
69-SM	&	35.6591	&	142.5	&	7.20459	&	4338.15 \\ 
116-SM	&	48.1897	&	50.625	&	7.14967	&	4253.01 \\ 
17-SM	&	17.6124	&	165	&	7.14067	&	6128.22 \\ 
166-SM	&	54.3147	&	247.5	&	7.10726	&	3184.32 \\ 
180-SM	&	60	&	50.625	&	6.88179	&	5757.5 \\ 
89-SM	&	41.8588	&	70.7143	&	6.83453	&	2849.03 \\ 
165-SM	&	54.3147	&	236.25	&	6.82174	&	3554.79 \\ 
135-SM	&	48.1897	&	264.375	&	6.80428	&	7532.09 \\ 
$\vdots$	&	$\vdots$	&	$\vdots$	&	$\vdots$	&	$\vdots$ \\ 
4-DM	&	11.7159	&	22.5	&	3.30532	&	3529.66 \\ 
151-SM	&	54.3147	&	78.75	&	3.26038	&	1616.3 \\ 
183-SM	&	60	&	84.375	&	3.25671	&	1092.14 \\ 
246-DM	&	70.5288	&	73.125	&	3.25244	&	1340.67 \\ 
36-DM	&	23.5565	&	281.25	&	3.24849	&	4173.87 \\ 
90-SM	&	41.8588	&	83.5714	&	3.22798	&	2345.33 \\ 
104-DM	&	41.8588	&	263.571	&	3.22568	&	2562.22 \\ 
90-DM	&	41.8588	&	83.5714	&	3.21068	&	2157.03 \\ 
120-DM	&	48.1897	&	95.625	&	3.19303	&	1828.14 \\ 
115-DM	&	48.1897	&	39.375	&	3.1624	&	3865.41 \\
24-DM	&	23.5565	&	11.25	&	3.15459	&	1065.64 \\ 
25-DM	&	23.5565	&	33.75	&	3.14698	&	3109.02 \\ 
27-DM	&	23.5565	&	78.75	&	3.11232	&	2617.48 \\ 
277-DM	&	75.5225	&	56.25	&	3.10658	&	3516.82 \\ 
309-SM	&	80.4059	&	61.875	&	3.08276	&	1562.32 \\ 
62-DM	&	35.6591	&	37.5	&	3.06531	&	3248.03 \\ 
23-SM	&	17.6124	&	345	&	3.06438	&	4181.3 \\ 
35-DM	&	23.5565	&	258.75	&	3.04779	&	6169.51 \\ 
196-SM	&	60	&	230.625	&	3.03069	&	4741.18 \\ 
65-DM	&	35.6591	&	82.5	&	3.02751	&	1904.96 \\ 
1-DM	&	5.85027	&	135	&	3.01031	&	5129.44 \\ 
\hline
\end{tabular}
\end{table}
\footnotetext{ The full list of sub-sample can be found in Footnote$^{~\ref{ft:video}}$.}

\begin{table}[!htbp]
\caption{Phase-resolved spectral fitting results }\label{table:resultstable}
\centering
\begin{tabular}{cccc}
\hline\hline
Phase range	&	Spectral Index	&	Normalization	&	reduced $\chi^2$ \\ 
\hline
0.00-0.01	&	2.137 $\pm$ 0.0264	&	10.370 $\pm$ 1.2108	&	1.069	\\
0.01-0.02	&	2.185 $\pm$ 0.0436	&	7.944 $\pm$ 1.5154	&	1.056	\\
0.02-0.03	&	2.094 $\pm$ 0.0554	&	4.000 $\pm$ 1.0002	&	1.047	\\
0.03-0.05	&	2.063 $\pm$ 0.0525	&	2.624 $\pm$ 0.6270	&	1.023	\\
0.05-0.07	&	2.073 $\pm$ 0.0688	&	2.086 $\pm$ 0.6259	&	1.014	\\
0.07-0.09	&	1.954 $\pm$ 0.0819	&	0.992 $\pm$ 0.3704	&	1.023	\\
0.09-0.11	&	2.080 $\pm$ 0.1059	&	1.405 $\pm$ 0.6457	&	1.016	\\
0.11-0.13	&	2.069 $\pm$ 0.0896	&	1.571 $\pm$ 0.6156	&	1.023	\\
0.13-0.15	&	2.035 $\pm$ 0.0946	&	1.264 $\pm$ 0.5301	&	1.024	\\
0.15-0.17	&	1.996 $\pm$ 0.0908	&	1.093 $\pm$ 0.4466	&	1.015	\\
0.17-0.19	&	2.019 $\pm$ 0.1067	&	1.037 $\pm$ 0.4957	&	1.020	\\
0.19-0.21	&	2.025 $\pm$ 0.0812	&	1.401 $\pm$ 0.5060	&	1.002	\\
0.21-0.23	&	2.044 $\pm$ 0.0788	&	1.587 $\pm$ 0.5521	&	1.024	\\
0.23-0.25	&	1.962 $\pm$ 0.0784	&	1.079 $\pm$ 0.3831	&	1.008	\\
0.25-0.27	&	2.015 $\pm$ 0.0594	&	1.828 $\pm$ 0.4855	&	1.046	\\
0.27-0.29	&	1.972 $\pm$ 0.0521	&	1.704 $\pm$ 0.4170	&	1.006	\\
0.29-0.31	&	1.991 $\pm$ 0.0482	&	2.024 $\pm$ 0.4565	&	1.032	\\
0.31-0.33	&	1.979 $\pm$ 0.0394	&	2.329 $\pm$ 0.4305	&	1.022	\\
0.33-0.35	&	1.997 $\pm$ 0.0323	&	3.099 $\pm$ 0.4661	&	1.025	\\
0.35-0.37	&	2.026 $\pm$ 0.0275	&	4.203 $\pm$ 0.5320	&	1.054	\\
0.37-0.38	&	2.036 $\pm$ 0.0324	&	5.160 $\pm$ 0.7693	&	1.054	\\
0.38-0.39	&	2.029 $\pm$ 0.0267	&	6.052 $\pm$ 0.7468	&	1.063	\\
0.39-0.40	&	1.997 $\pm$ 0.0243	&	5.700 $\pm$ 0.6452	&	1.097	\\
0.40-0.41	&	1.994 $\pm$ 0.0260	&	5.264 $\pm$ 0.6377	&	1.066	\\
0.41-0.42	&	2.098 $\pm$ 0.0384	&	5.878 $\pm$ 1.0181	&	1.076	\\
0.42-0.43	&	2.101 $\pm$ 0.0596	&	3.814 $\pm$ 0.9885	&	1.049	\\
0.43-0.44	&	2.250 $\pm$ 0.1026	&	4.593 $\pm$ 1.9297	&	1.050	\\
0.44-0.46	&	2.091 $\pm$ 0.1042	&	1.502 $\pm$ 0.6825	&	1.015	\\
0.46-0.48	&	2.223 $\pm$ 0.2689	&	0.930 $\pm$ 0.6810	&	1.018	\\
0.48-0.52	&	2.317 $\pm$ 0.2781	&	1.288 $\pm$ 1.5232	&	1.019	\\
0.52-0.60	&	1.937 $\pm$ 0.4959	&	0.086 $\pm$ 0.2019	&	1.001	\\
0.80-0.88	&	1.718 $\pm$ 0.3187	&	0.048 $\pm$ 0.0724	&	1.016	\\
0.88-0.92	&	2.167 $\pm$ 0.1846	&	0.920 $\pm$ 0.7491	&	1.013	\\
0.92-0.94	&	2.238 $\pm$ 0.1176	&	2.734 $\pm$ 1.3221	&	0.995	\\
0.94-0.95	&	2.315 $\pm$ 0.1425	&	4.587 $\pm$ 2.6947	&	1.030	\\
0.95-0.96	&	2.258 $\pm$ 0.0956	&	5.130 $\pm$ 1.9946	&	1.039	\\
0.96-0.97	&	2.299 $\pm$ 0.0721	&	8.314 $\pm$ 2.4149	&	1.070	\\
0.97-0.98	&	2.221 $\pm$ 0.0490	&	8.432 $\pm$ 1.7850	&	1.047	\\
0.98-0.99	&	2.174 $\pm$ 0.0352	&	9.276 $\pm$ 1.4318	&	1.090	\\
0.99-1.00	&	2.124 $\pm$ 0.0246	&	10.421 $\pm$ 1.1298	&	1.102	\\
\hline
\end{tabular}
\end{table}

\end{document}